\documentclass[11pt,a4paper]{article}

\usepackage{times,epsfig}

\advance \textwidth by -5cm
\advance \textheight by -6cm
\def\e{\begin{equation}}
\def\f{\end{equation}}
\def\_#1{{\bf #1}}

\def\.{\cdot}

\begin{document}

\title{\bfseries{Unity absorbance layers - optimal design criteria}}

\def\affil#1{\begin{itemize} \item[] #1 \end{itemize}}

\author{\bfseries K. Chalapat$^1$, G. S. Paraoanu$^1$,  Z. Du$^2$, J. Tervo$^2$, I. Nefedov$^2$, S. Tretyakov$^2$}



\date{}

\maketitle
\thispagestyle{empty}

\affil{$^1$Low Temperature Laboratory, Aalto University, P.O. Box 15100, FI-00076 Aalto, Finland\\
$^2$Department of Radio Science and Engineering, Aalto University, P.O. Box 13000, FI-00076, Aalto, Finland}

\begin{abstract}

\noindent We present the necessary and sufficient  conditions for
the unity absorbance of thin planar layers. With a simple structure
comprising a double-layer grid, it is shown that zero transmission
and reflection is feasible only if the second sheet is a perfectly
conducting wall. The operational frequencies can be varied by tuning
the grid impedance. Explicit conditions are given at the large
wavelength limit; these underline the fundamental of ultra-thin
perfect antennae and absorbers.
\end{abstract}

\subsection*{1.~Introduction}

In this paper we will discuss the possibility  to design and
realize planar layers that absorb all incident power of plane waves
at normal incidence (zero reflection, zero transmission, unity
absorbance). It is obvious that if the layer is an infinitely thin
sheet of induced electric current, this is not possible to realize.
This is because a sheet of electric current radiates symmetrically
in the forward and back direction, so the requirement of zero
reflection means the absence of shadow (unity transmission). Thus,
if no magnetic materials are to be used, the unity-absorbance layer
must have non-zero thickness. If the back surface is an impenetrable wall, for example a perfect electric conductor (PEC) surface, the design of unity absorbance layers is trivial, \cite{Fante}, for example, by positioning a resistive sheet at the quarter-wave distance from the PEC plane and choose its electrical surface admittance to be equal to that of free space. The sheet can be brought closer to the wall by adding appropriate reactance in the
sheet response (such absorbers are well known). We are, however,
interested in the case when no impenetrable wall is present. This
way one can realize layers that are transparent in some frequency
ranges but absorb all power at some other frequencies. Recently,
some numerical and experimental results for such absorbing layers
were published, but the design was based on numerical simulations,
and no ideal performance (zero transmission and zero reflection at
the same frequency) has been demonstrated. Here, we will
theoretically consider a possible realization using two parallel
infinitely thin sheet separated by a free-space layer.

\subsection*{2.~Matched unity absorbance layers criteria}

Natural materials are known to exhibit high absorption at some
frequencies. But tailoring a natural material to achieve perfect
absorption over a frequency bandwidth is not straightforward. Within
the last two years, there has been many  attempts to realize devices
with near unity absorbance \cite{Kravets}-\cite{Wang}.

\begin{figure}
  \centering
  \includegraphics[scale=0.5]{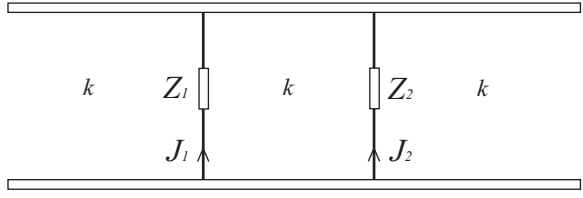}
  \caption{The transmission line model of an antenna composed of two planar grids. \label{TransmissionLineModel}}
\end{figure}

In this paper, we discuss and analyze the optimal criteria  to
realize perfect absorption at normal incidence. The theory is drawn
based on the transmission-line model of a double-layer planar grids,
Fig.~\ref{TransmissionLineModel}. The structure is specified with
grid impedances $Z_j$ and average surface current densities $J_j$.
By applying the the grid condition, $E^{\rm tot}_x = Z_g J$, \cite{Tretyakov}, we obtain, at the first load,
\begin{equation}
E^{\rm inc}-\frac{\eta J_1}{2} -\frac{\eta J_2}{2} e^{-jkd} = Z_1 J_1,
\label{bc1}
\end{equation}
and at the second load,
\begin{equation}
E^{\rm inc} e^{-jkd}-\frac{\eta J_1}{2} e^{-jkd} -\frac{\eta J_2}{2} = Z_2 J_2.
\label{bc2}
\end{equation}

Solving Eq.~(\ref{bc1}) and (\ref{bc2}), we determine the average
current densities.
The reflected field $E^{\rm ref}$ and transmitted $E^{\rm tran}$
fields are related to the current densities on the grids as
\begin{equation}
E^{\rm ref} = -\frac{\eta}{2} J_1 - \frac{\eta}{2} J_2 e^{-jkd},
\label{Eref}
\end{equation}
\begin{equation}
E^{\rm tran} = E^{\rm inc}e^{-jkd} -\frac{\eta}{2} J_1 e^{-j k d} -
\frac{\eta}{2} J_2. \label{Etran}
\end{equation}
By definition, $R = E^{\rm ref}/E^{\rm inc}$ and $T = E^{\rm
tran}/(E^{\rm inc} e^{-j k d})$, and we can express the reflection
and transmission coefficients in terms of the grid parameters:
\begin{equation}
R = -\frac{ \frac{\eta}{2} (Z_2 + Z_1 e^{-j2kd}) + (\frac{\eta}{2})^2 (1-e^{-j 2kd}) }{\frac{\eta}{2}Z_1 + \frac{\eta}{2}Z_2+ Z_1Z_2 + (\frac{\eta}{2})^2(1-e^{-j 2kd})},
\label{R}
\end{equation}
\begin{equation}
T = \frac{Z_1 Z_2}{\frac{\eta}{2}Z_1 + \frac{\eta}{2}Z_2+ Z_1Z_2 + (\frac{\eta}{2})^2(1-e^{-j 2kd})}.
\label{T}
\end{equation}

The trivial solutions for zero transmission are the cases in which one of the grid impedances is zero, $i.e.$ a short-circuit.
To obtain non-trivial solutions for total absorption, we firstly set the reflection coefficient in Eq.~(\ref{R}) to zero to obtain the optimal condition for the grid impedances
\begin{equation}
Z_2 = -Z_1 e^{-j 2kd} - \frac{\eta}{2} \left( 1 - e^{-j2kd} \right).
\label{Z2Z1R0}
\end{equation}
Substituting this back to Eq.~(\ref{T}) reveals the relation between the transmission coefficient and $Z_1$:
\begin{equation}
T = 1+\frac{\eta}{2 Z_1} \left( e^{j 2kd}-1 \right).
\label{T_Z1}
\end{equation}
Eqs.~(\ref{Z2Z1R0}) and (\ref{T_Z1}) constitute the conditions for
unity absorbance  in terms of the grid parameters and electrical
thickness of the spacing: $Z_1$, $d$ and $k$. Note that $k$ is the
wave vector of the medium filling the space between the two grids,
which is assumed to the same as the surrounding medium.

Equating the transmission coefficient (\ref{T_Z1}) to zero, we find
the required impedance of the first sheet:
\begin{equation}
Z_1={\eta \over 2}\left(1-e^{2jkd}\right). \label{Z1R0T0}
\end{equation}
Finally, substituting into (\ref{Z2Z1R0}) we find that $Z_2=0$.


\subsection*{3.~Discussion and conclusion}

The results show that it is actually not possible to realize perfect absorbers (such that at some frequency both $R=T=0$) except if the second sheet is a perfectly conducting layer. The well-known absorber design based on a lossy and possibly reactive thin sheet parallel to a PEC wall is seen to be the only solution. For thin layers, using the Taylor expansion of Eq. (\ref{Z1R0T0}), the required grid impedance of the first sheet reads
\begin{equation}
Z_1\approx \eta (kd)^2 - j\eta k d.
\end{equation}
If only passive materials are used in the design, the ideal operation is possible only at a single frequency. This is obvious since the required reactance of the first grid corresponds to a \emph{lossless negative inductance} and the frequency-decreasing reactance of a lossless structure is incompatible with the causality principle of material response (the Kramers-Kronig relations and the Foster theorem are violated). Wide bandwidth in this design is possible only if active or tunable components are used. Another classical design follows, from Eq.  (\ref{Z1R0T0}), when the first grid is a quarter wavelength away from the PEC wall ($d = \lambda/4$), in this case the required $Z_1$ is purely resistive with $Z_1=\eta$.

In the future, we plan to study possibilities offered by more
general configurations, with material layers between the grids or
more than two grids, and find out if unit-absorbance layers can be
designed without utilizing perfect electric conductors.

\subsubsection*{Acknowledgements}
This topic was first proposed by ST as a subject of a student
research project within the post-graduate course ``Analytical
modeling in applied electromagnetics'' (Helsinki University of
Technology, 2009). The work is financially supported by Thailand Commission on Higher Education and the Academy of Finland.


{\small


\begin{thebibliography}{10}
\setlength{\itemsep}{-1ex}

\bibitem{Fante} R. L. Fante and M. T. McCormack, Reflection properties of the salisbury screen, \emph{IEEE. Trans. Antennas Propagat.,} vol. 36, p. 1443-1454, 1988.

\bibitem{Kravets} V. G. Kravets, F. Schedin, and A. N. Grigorenko, Plasmonic
blackbody: Almost complete absorption of light in nanostructured
metallic coatings, \emph{Phys. Rev. B.,} vol. 78, p. 205405, 2008.

\bibitem{Landy} N. I. Landy, S. Sajuyigbe, J. J. Mock, D. R. Smith, andW. J.
Padilla, Perfect metamaterial absorber, \emph{Phys. Rev. Lett.,}
vol. 100, p. 207402, 2008.

\bibitem{Tao} H. Tao, N. I. Landy, C. M. Bingham, X. Zhang, R. D. Averitt,
and W. J. Padilla, A metamaterial absorber for the terahertz regime:
design, fabrication and characterization, \emph{Optics Express,}
vol. 16, p. 7181, 2008.

\bibitem{Zhu} Zhu Bo, Wang Zheng-Bin, Yu Zhen-Zhong, Zhang Qi, Zhao
Jun-Ming, Feng Yi-Jun, Jiang Tian, Planar metamaterial microwave
absorber for all wave polarizations, \emph{Chin. Phys. Lett.,} vol.
26, 114102, 2009.

\bibitem{Diem} M. Diem, T. Koschny, and C. M. Soukoulis, Wide-angle perfect
absorber/thermal emitter in the terahertz regime, \emph{Phys. Rev.
B,} vol. 79, 033101, 2009.

\bibitem{Driessen1} E. F. C. Driessen and M. J. A. de Dood, The perfect absorber,
\emph{Appl. Phys. Lett.,} vol. 94, 171109, 2009.

\bibitem{Wang} J. F. Wang, S. B. Qu, Z. T. Fu, H. Ma, Y. M. Yang and X. Wu,
Three-dimensional metamaterial microwave absorbers composed of
coplanar magnetic and electric resonators, \emph{Progress in
Electromagnetics Research Letters,} vol. 7, p. 15-24, 2009.

\bibitem{Tretyakov} S. Tretyakov, \emph{Analytical Modeling in Applied
Electromagnetics,} Artech House, 2003.


\end{thebibliography}
}


\end{document}